# A Novel Adaptive Channel Equalization Method Using Variable Step-Size Partial Rank Algorithm

Sayed A. Hadei                    Paeiz Azmi

Faculty of Electrical and Computer Engineering,Tarbiat Modares University,
P.O.Box 14115-143, Tehran, IRAN
E-mail: {a.hadei, pazmi}@modares.ac.ir

*Abstract*— **Recently a framework has been introduced within which a large number of classical and modern adaptive filter algorithms can be viewed as special cases. Variable Step-Size (VSS) normalized least mean square (VSSNLMS) and VSS Affine Projection Algorithms (VSSAPA) are two particular examples of the adaptive algorithms that can be covered by this generic adaptive filter. In this paper, we introduce a new VSS Partial Rank (VSSPR) adaptive algorithm based on the generic VSS adaptive filter and use it for channel equalization. The proposed algorithm performs very well in attenuating noise and inter-symbol interference (ISI) in comparison with the standard NLMS and the recently introduced AP algorithms.**

## I. Introduction

In modern digital communications, it is well known that channel equalization plays an important role in compensating channel distortion. Unfortunately, most channels have time-varying characteristic and their transfer functions change with time. Furthermore, time-varying multipath interference and multiuser interference are two major limitations for high speed digital communications. Usually, adaptive equalizers are applied in order to cope with these issues [1].

For adaptive channel equalization, we need a suitable filter structure and proper adaptive algorithms. High-speed digital transmissions mostly suffer from inter-symbol interference (ISI) and additive noise. The adaptive equalization algorithms recursively determine the filter coefficients in order to eliminate the effects of noise and ISI [2]-[4].

Among the numerous algorithms that can be used for adaptive filtering, the Least Mean Square (LMS) algorithm has enjoyed widespread popularity because of its simplicity in computation and implementation. However, it is well known that the LMS type algorithms can only minimize the current estimate error to some extent. It is known that a variable step-size algorithm has to be applied to make a trade-off between the convergence rate and the steady-state misadjustment. On the other hand, the best adaptive filter algorithm, i.e., Recursive Least Square (RLS) converges significantly faster than the LMS algorithm, but its computational complexity and numerical instability may prohibit its usage in many applications [5]. To solve this problem, a number of affine subspace projections-based adaptive filtering structures, such as the standard version of affine projection algorithm (APA), the regularized APA (R-APA), the partial rank algorithm (PRA) [6], and multirate techniques, have been proposed in the literature [7]. In these algorithms, a properly selected fixed step size can change the convergence and the steady state mean square error. By optimally selecting the step size during the adaptation, we can obtain both fast convergence rate and low steady state mean square error. Variable Step Size (VSS) versions of the NLMS and the APA algorithm (APA) are two important examples of these algorithms [8].

Our objective in this paper is firstly to develop a VSS version of the generic adaptive filter of [9], which was shown to cover many classical and modern adaptive filter algorithms as special cases. We proceed in the following by presenting a VSS version of partial rank algorithm, which is called VSSPR adaptive filter. The proposed filter is characterized by its fast convergence speed, and reduced steady state mean square error in comparison with the ordinary PRA. Then we use it for channel equalization.

We have organized our paper as follows. In the following section, the generic variable step size update equation forming the basis of our development of the VSSPR is introduced. Subsequently, the VSSPR adaptive algorithm will be presented in Section III. In Sections IV and V, we show the VSSPR algorithm performance in adaptive channel equalization usage for attenuating noise and inter-symbol interference and the results will be compared to the NLMS, AP, and PR algorithms. Section VI concludes this paper.

## II. The Generic Variable Step-Size Update Equation

Throughout this paper, we use a notation based on the adaptive filtering setup that is shown in Fig. 1 and explained in Table I. The generic filter vector update equation, which is at the center of our analysis, can be stated as follow [9]:

$$\underline{h}(n+1) = \underline{h}(n) + \mu\, X(n) W(n) \underline{e}(n). \qquad (1)$$

where the error signal vector $\underline{e}(n)$ is calculated as [5]:

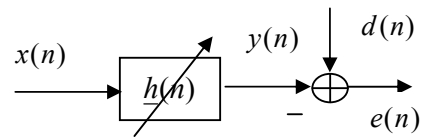

Figure 1: A prototypical adaptive filter setup

TABLE I. EXPLANATION OF NOTATIONS

| | |
|---|---|
| $\underline{h}(n)$ | Length-$M$ column vector of filter coefficients to be adjusted at each time instant $n$ |
| $\underline{x}(n)$ | Length-$M$ vector of input signal samples to adaptive filter, $[x(n), x(n-1), \ldots, x(n-M+1)]^T$ |
| $\underline{e}(n)$ | Length-$L$ vector of error samples. $[e(n), e(n-1), \ldots, e(n-L+1)]^T$ |
| $X(n)$ | $M \times L$ signal matrix whose columns are given by $\underline{x}(n), \underline{x}(n-1), \ldots, \underline{x}(n-L+1)$ |
| $W(n)$ | $L \times L$ symmetric weighting matrix |
| $\mu$ | Step-size |

TABLE II. CORRESPONDENCE BETWEEN SPECIAL CASES OF EQ. 1 AND VARIOUS ADAPTIVE FILTERING ALGORITHMS

| $L$ | $W(n)$ | Algorithm |
|---|---|---|
| 1 | 1 | LMS |
| 1 | $\|\underline{x}(n)\|^{-2}$ | NLMS |
| $1 < L < M$ | $[X^T(n)X(n)]^{-1}$ | APA |
| $1 < L < M$ | $[\varepsilon I + X^T(n)X(n)]^{-1}$ | R-APA |

$$\underline{e}(n) = \underline{d}(n) - X^T(n)\underline{h}(n) \qquad (2)$$

Based on Eq. 1 and by specific choosing $L$ and $W(n)$ several adaptive filter algorithms such as LMS, NLMS, and AP algorithms can be derived [9]. The particular choices and their corresponding algorithms are summarized in Table II.

Eq. 2 can be stated in terms of weight error vector, $\underline{\in}(n) = \underline{h}_t - \underline{h}(n)$, where $\underline{h}_t$ is the unknown true filter vector [10, p.91, Eq.4]:

$$\underline{\in}(n+1) = \underline{\in}(n) - \mu X(n)W(n)\underline{e}(n), \qquad (3)$$

Taking, the squared norm and expectations from both sides of Eq. 3 we have,

$$\begin{aligned}
E\left\{\|\underline{\varepsilon}(n+1)\|^2\right\} &= E\left\{\|\underline{\in}(n)\|^2\right\} + \\
\mu^2 E\left\{\underline{e}^T(n)B^T(n)B(n)\underline{e}(n)\right\} &- \\
2\mu E\left\{\underline{e}^T(n)B^T(n)\underline{\in}(n)\right\}
\end{aligned} \qquad (4)$$

where $B(n) = X(n)W(n)$. Eq. 4 can be represented as follows:

$$E\left\{\|\underline{\in}(n+1)\|^2\right\} = E\left\{\|\underline{\in}(n)\|^2\right\} - \Delta\mu \qquad (5)$$

where $\Delta\mu$ is:

$$\Delta\mu = -\mu^2 E\left\{\underline{e}^T(n)B^T(n)B(n)\underline{e}(n)\right\} + 2\mu E\left\{\underline{e}^T(n)B^T(n)\underline{\in}(n)\right\}, \qquad (6)$$

If $\Delta\mu$ is maximized, then mean-square deviation (MSD) will undergo the largest decrease from iteration $n$ to iteration $n+1$. So, the optimum step-size $\mu^o(n)$ will be found as,

$$\mu^0(n) = \frac{E\left\{\underline{e}^T(n)B^T(n)\underline{\in}(n)\right\}}{E\left\{\underline{e}^T(n)B^T(n)B(n)\underline{e}(n)\right\}}. \qquad (7)$$

Introducing a-prior-error vector as follows [10, p.92]:

$$\underline{e}_a(n) = X^T(n)\underline{\in}(n), \qquad (8)$$

we find that the error vectors are related to the a-priori-error vectors as follow:

$$\underline{e}(n) = \underline{e}_a(n) + \underline{v}(n) \qquad (9)$$

Assuming the noise sequence $v(n)$ is identically and independently distributed and statistically independent of the regression data, and neglecting the dependency of $\in(n)$ on the past noises, we establish the following two sub equations:

Part I:

$$\begin{aligned}
&E\left\{\underline{e}^T(n)B^T(n)\underline{\in}(n)\right\} = \\
&E\left\{\left(\in^T(n)X(n) + \underline{v}^T(n)\right)\left(B^T(n)\underline{\in}(n)\right)\right\} = \\
&E\left\{\underline{\in}^T(n)X(n)B^T(n)\underline{\in}(n)\right\}
\end{aligned} \qquad (10)$$

Part II:

$$\begin{aligned}
&E\left\{\underline{e}^T(n)B^T(n)B(n)\underline{e}(n)\right\} = \\
&E\left\{\left(\in^T(n)X(n)B^T(n)B(n)X^T(n)\underline{\in}(n)\right)\right\} + \\
&E\left\{\underline{v}^T(n)B^T(n)B(n)\underline{v}(n)\right\} = \\
&E\left\{\underline{\in}^T(n)X(n)B^T(n)B(n)X^T(n)\underline{\in}(n)\right\} + \\
&\sigma_v^2 Tr\left(E\left\{B^T(n)B(n)\right\}\right)
\end{aligned} \qquad (11)$$

Finally, with defining $C(n) = B(n)X^T(n)$, the optimum step-size in Eq. 11 becomes:

$$\mu^o(n) = \frac{E\left\{\underline{\in}^T(n)C^T(n)\underline{\in}(n)\right\}}{E\left\{\underline{\in}^T(n)C^T(n)C(n)\underline{\in}(n)\right\} + \psi} \qquad (12)$$

where

$$\psi = \sigma_v^2 Tr\left(E\left\{B^T(n)B(n)\right\}\right) \qquad (13)$$

Substituting the $\mu^o(n)$ of Eq. 12 instead of $\mu$ in Eq. 1 the general variable step-size update equation that covers VSSNLMS, VSSAPA, and other VSS algorithms as special cases, will be obtained. We now focus on the development of the VSSPR adaptive filter.

### III. The Variable Step-Size Partial Rank Adaptive Filter Algorithm

The vector update equation for PRA can be stated as [6]:

$$\underline{h}(n+1) = \underline{h}(n-L') + \mu X(n) \left( \in I + X^T(n)X(n) \right)^{-1} \underline{e}(n), \quad (14)$$

where $L' = \alpha(L-1)$. Selecting $\alpha = 0$ results in affine projection algorithm. When $\alpha = 1$, the partial rank (PR) adaptive filter algorithm is obtained and it means that the weight vector is updated only once after $L$ iterations. From Eq.12, the optimum step size can be found as:

$$\mu^o(n) = \frac{E\left\{ P(n) \right\}}{E\left\{ Q(n) \right\} + \psi} \quad (15)$$

where,

$$P(n) = \underline{e}^T(n) \, X(n) \left( \in I + X^T(n)X(n) \right)^{-1} X^T(n) \underline{e}(n), \quad (16)$$

and,

$$\begin{aligned} Q(n) = &\underline{e}^T(n) \, X(n) \left( \in I + X^T(n)X(n) \right)^{-1} X^T(n) \\ &X(n)\left( \in I + X^T(n)X(n) \right)^{-1} X^T(n)\underline{e}(n) \end{aligned} \quad (17)$$

by defining,

$$p(n) = X(n) \left( \in I + X^T(n)X(n) \right)^{-1} X^T(n)\underline{e}(n), \quad (18)$$

and also by using the following approximation, [8, P.134]:

$$\left( \in I + X^T(n)X(n) \right)^{-1} X^T(n)X(n) \approx I \quad (19)$$

we obtain,

$$\mu^o(n) = \frac{E\left\{ \left\| \underline{p}(n) \right\|^2 \right\}}{E\left\{ \left\| \underline{p}(n) \right\|^2 \right\} + \psi} \quad (20)$$

where $\| \, . \, \|$ represents the squared Euclidean norm of a vector and $\Psi$ is a positive constant and can be found from:

$$\psi = \sigma_\upsilon^2 \, Tr\left( \, E\left\{ \left( \in I + X^T(n)X(n) \right)^{-1} \right\} \right) \quad (21)$$

In calculating the optimum step size from Eq.20, the main problem is that $\underline{p}(n)$ is not available; this is because $\underline{h}_t$ is unknown. Therefore we need to estimate this quantity. Taking expectation from both side of Eq.18 we have:

$$E\left\{ \underline{p}(n) \right\} = E\left\{ X(n)(\in I + X^T(n)X(n))^{-1} X^T(n)\underline{e}(n) \right\} \quad (22)$$

and substituting $\underline{e}_a(n) = \underline{e}(n) - \underline{v}(n)$ into Eq.22 yields:

$$E\left\{ \underline{p}(n) \right\} = E\left\{ X(n)(\in I + X^T(n)X(n))^{-1} X^T(n)\underline{e}(n) \right\} \quad (23)$$

We can estimate this quantity with the recursion presented in Eq.24 as follow:

$$\underline{\hat{p}}(n) = \beta \underline{\hat{p}}([n-1]) + (1-\beta)X(n)(\in I + X^T(n)X(n))^{-1} \underline{e}(n) \quad (24)$$

where $\beta$ is a smoothing factor and $\left( 0 \le \beta \le 1 \right)$. Finally the recursion For Variable Step Size Partial Rank Adaptive Filter (VSSPR) is given by:

$$\underline{h}(n+1) = \underline{h}(n-L') + \mu(n) X(n) \left( \in I + X^T(n)X(n) \right)^{-1} \underline{e}(n), \quad (25)$$

$$\mu(n) = \mu_{max} \frac{\left\| \underline{\hat{p}}(n) \right\|^2}{\left\| \underline{\hat{p}}(n) \right\|^2 + \psi} \quad (26)$$

The step-size changes with the $\left\| \underline{\hat{p}}(n) \right\|^2$, and the constant $\Psi$, as it can be seen in Appendix I, can be approximated as $L/SNR$. To guarantee update stability, $\mu_{max}$ is selected less than 2 [10]. The final results of VSSPR adaptive filter algorithm have been summarized in Table III.

### IV. Adaptive Channel Equalizer

The application considered in this paper for adaptive filters is linear channel equalization. The block diagram a communication system with an adaptive equalizer, is shown in Fig. 2, in which data symbols $\{s(.)\}$ are transmitted through a channel and the output sequence is measured in the presence of additive noise $v(.)$. The signals $\{v(.), s(.)\}$ are assumed to be uncorrelated. The noisy output of the channel is denoted by $u(.)$ and is fed into a linear equalizer with $M$ taps. At any particular time instant $n$, the state of the equalizer is given by $\underline{x}(n) = [x(n), x(n-1), ...., x(n-M+1)]$.

It is desired to determine the equalizer tap vector $W$ in order to estimate the signal $d(n) = s(n-\Delta)$ optimally in the least-mean-square sense. At each time instant $n$, the symbol $d(n) = s(n-\Delta)$ is compared with the output of the adaptive filter $\hat{s}(n-\Delta)$ and an error signal $e(n) = d(n) - \underline{x}(n)\underline{w}(n-1)$ is generated. The error is then used to adjust the filter coefficient from $\underline{w}(n-i)$ to $\underline{w}(n)$ by using an algorithm like LMS or NLMS.

TABLE III. VSS PARTIAL RANK ADAPTIVE FILTER ALGORITHM

$$X(n) = \left[\underline{x}(n), \underline{x}(n-1), ..., \underline{x}(n-L+1)\right]$$

$$\hat{\underline{p}}(n) = \beta \, \hat{\underline{p}}([n-1]) +$$

$$(1-\beta)X(n)(\in I + X^T(n)X(n))^{-1}\underline{e}(n)$$

$$\mu(n) = \mu_{\max} \frac{\left\|\hat{\underline{p}}(n)\right\|^2}{\left\|\hat{\underline{p}}(n)\right\|^2 + \psi}, \quad \psi = L / SNR$$

$$\underline{h}(n+1) = \underline{h}(n-L') + \mu(n)X(n)\left(\in I + X^T(n)X(n)\right)^{-1}\underline{e}(n),$$

In this paper, we propose to adjust these filter coefficients with VSSPR algorithm. In steady state case, the error signal will have small values and, hence, the output $\hat{s}(n-\Delta)$ of the adaptive filter will get values close to $s(n-\Delta)$. It should be noted that this scheme for adaptive channel equalization does not require knowledge of the channel. In practice, following a training phase with a known reference sequence $\{d(n)\}$, an equalizer could continue to operate in one of two modes. In the first mode, its coefficient vector $\underline{w}$ would be frozen and used thereafter to generate future outputs $\{\hat{s}(n-\Delta)\}$. This mode of operation is appropriate when the training phase is successful enough to result in a reliable estimate $\hat{s}(n-\Delta)$, which is an estimate that lead to a low probability of error after feeding into a decision device mapping $\hat{s}(n-\Delta)$ to the closet point in the symbol constellation $\bar{s}(n-\Delta)$. However, if the channel varies slowly with time, it may be necessary that the equalizer continue to operate in a decision-directed mode. In this mode, the weight vector of the equalizer continues to be adapted even when the training phase has ended.

## V. EXPERIMENTAL RESULTS

In computer simulations, we consider a channel with the transfer function $C(z) = 0.5 + 1.2Z^{-1} + 1.5z^{-2} - z^{-3}$ and proceed to design an adaptive linear equalizer for it. As it is seen in figure 2, symbols $\{s(n)\}$ are transmitted through the channel and corrupted by additive complex valued white noise $\{v(n)\}$. The received signal $\{x(n)\}$ is processed by the FIR equalizer to generate estimates $\hat{s}(n-\Delta)$, which are fed into a decision device. The equalizer processes in two modes of operation: a training mode during which a delayed replica of the input sequence is used as a reference sequence, and a decision-directed mode during which the output of the decision-device replaces with the reference sequence. In the training method, a signification fraction of channel bandwidth is wasted to do the training action. Therefore, we must optimize the training signal to minimize the performance loss caused by channel estimation.

In the simulations, the input sequence $\{s(n)\}$ is chosen from a QPSK. The adaptive filter is trained with 500 symbols from a QPSK, followed by decision-directed operation during 5000 symbols from a 256-QAM constellation.

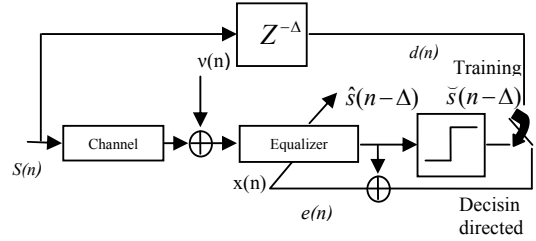

Figure 2: Adaptive linear equalizer operating in two modes: training mode and decision-directed mode.

The noise variance $\sigma_v^2$ has been chosen in order to enforce an SNR level of 30dB at the input of the equalizer. We have selected $\Delta=15$, $M=35$. The NLMS, PR, AP and VSSPR algorithms have been used to train the equalizer with step size $\mu=0.4$ for NLMS, $\mu=0.4$, L=4 for PR, $\mu=0.06$, L=4 for AP and $\mu_{max}=1.7$, $\psi=0.0001$, and L=4 for the VSSPR algorithm. Figure 3 show the learning curves. The simulated learning curves were obtained over 300 independent realizations and show the convergence of the VSSPRA has faster rate than other algorithms. Figure 4 shows symbol error rate (SER) curves versus signal to noise ratio at the input of the equalizer. The SER for the proposed algorithm is substantially lower than those of the other algorithms. We assumed that the receiver has knowledge of the transmitted information sequence informing the error signal between the desired symbol and its estimate. Such knowledge can be made available during a short training period in which a signal with a known information sequence is transmitted to the receiver for initially adjusting the tap weights. Figure 5 shows the training sequence and transmitted sequence. Figure 6 shows scatter diagrams of the received sequence and equalizer output using NLMS, PRA, APA and VSSPRA with 500 training symbols. The results show that the output of the equalizer with VSSPR algorithm is better than the other algorithms.

## VI. CONCLUSIONS

In this paper, we have developed a new Variable Step Size Partial Rank (VSSPR) algorithm based on unified framework of [9]. An application of this algorithm in channel equalization has been presented and its performance has been compared with those of some other algorithms. The proposed algorithm has good convergence rate and lower steady state mean square error in comparison with the ordinary PR, NLMS and APA algorithms.

## ACKNOWLEDGMENT

The authors are very grateful to the anonymous reviewers for their very insightful comments. This work is supported in

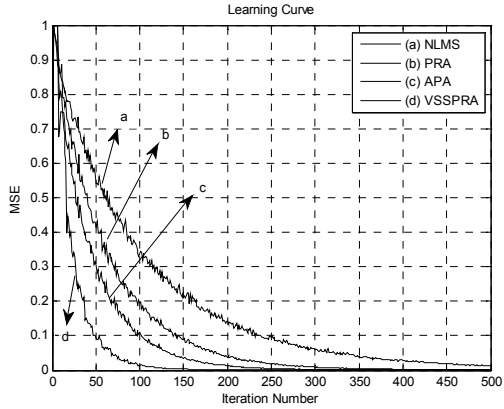

Figure 3: Learning Curve for the NLMS, PR, AP and VSSPR Algorithms with (μ=0.4 for NLMS, μ=0.4, L=4 for PR, μ=0.06, L=4 for AP and μmax=1.7, $\psi = 0.0001$, L=4 for the VSSPR).

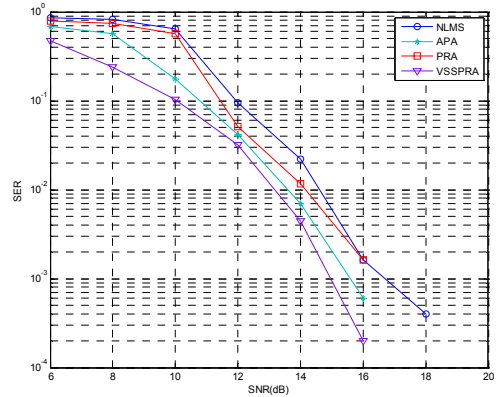

(a)

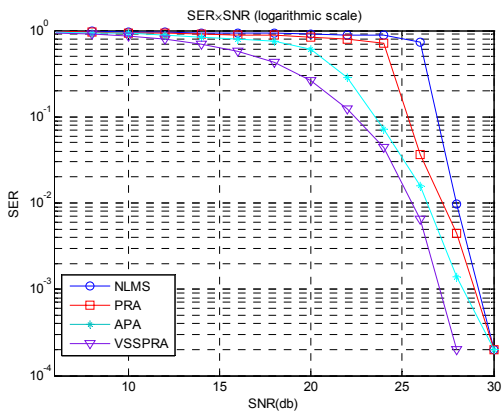

(b)

Figure 4: A plot of the SER as a function of the SNR at the input of the equalizer and the order of the QAM constellation ((a) 16-QAM, (b) 256-QAM). The adaptive filter is trained with NLMS, PRA, APA, and VSSPRA using 500 QPSK training symbols.

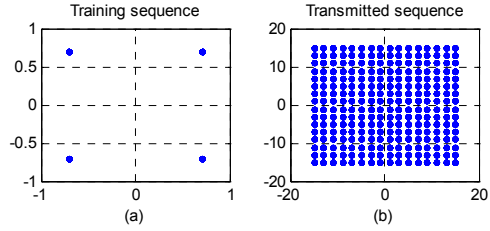

Figure 5: (a) scatter diagram of the Q-PSK training sequence. (b) Transmitted data.

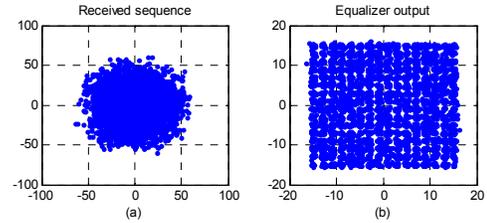

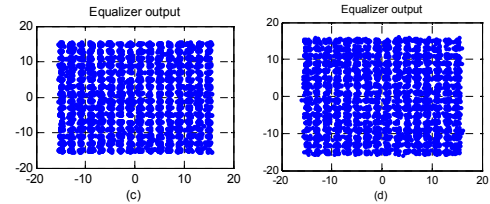

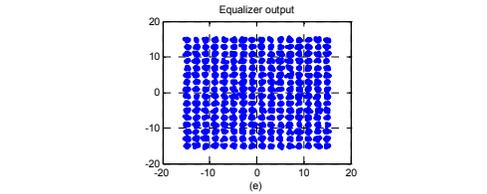

Figure 6: (a) Scatter diagram of the received sequence. (b) The output of the equalizer through the NLMS algorithm. (c) The output of the equalizer through the PR algorithm. (d) The output of the equalizer through the AP algorithm. (e) The output of the equalizer through the VSSPR algorithm.

# APPENDIX I

### FINDING AN APPROXIMATION FOR $\psi$

The positive constant $\psi$ is related to $\psi = \sigma_v^2 Tr(E\left\{(\in I + X^T(n)X(n))^{-1}\right\})$ . Using the assumption that $\in$ is small, the $\psi$ quantity can be found as follows:

$$\psi = \sigma_v^2 Tr(E\left(\begin{pmatrix} x(n) & x(n-1) & \dots & x(n-M+1) \\ x(n-1) & x(n-2) & \ddots & x(n-M) \\ \vdots & \vdots & \ddots & \vdots \\ x(n-L+1) & x(n-L) & & x(n-L-M+2) \end{pmatrix} \\ \times \begin{pmatrix} x(n) & x(n-1) & \dots & x(n-L+1) \\ x(n-1) & x(n-2) & \dots & x(n-L) \\ \vdots & \vdots & \ddots & \vdots \\ x(n-M+1) & x(n-M) & \dots & x(n-L-M+2) \end{pmatrix}\right)^{-1})$$

$$(27)$$

To approximate the Eq.27, we assume that the sequence $\underline{x}(n)$ is independent and identically distributed. Using this assumption, we can assume that $X^T(n)X(n)$ is close to diagonal. Therefore we focused just on the diagonal elements of Eq.27. Simplifying this equation, we obtain

$$\psi = \sigma_v^2 Tr(E\begin{pmatrix} \frac{1}{\|\underline{x}(n)\|^2} & \times & \dots & \times \\ \times & \frac{1}{\|\underline{x}(n-1)\|^2} & \dots & \times \\ \vdots & \vdots & \ddots & \vdots \\ \times & \times & \dots & \frac{1}{\|\underline{x}(n-L+1)\|^2} \end{pmatrix})$$

$$(28)$$

Applying the expectation and trace operators, we obtain

$$\psi = \sigma_v^2 Tr(E\left\{\frac{1}{\|\underline{x}(n)\|^2}\right\} + E\left\{\frac{1}{\|\underline{x}(n-1)\|^2}\right\} + \dots + E\left\{\frac{1}{\|\underline{x}(n-L+1)\|^2}\right\})$$

$$(29)$$

Eq.30 can be stated as :

$$\psi = L.\sigma_v^2 r E\left\{\frac{1}{\|\underline{x}(n)\|^2}\right\}$$

$$(30)$$

Now from [8], we obtain that $\psi$ can be approximated as $L/SNR$ . Therefore $\psi$ is inversely proportional to *SNR* and proportional to L.